\documentclass[preprint]{aastex}

\slugcomment{Draft for the Astrophysical Journal, \today}
\shorttitle{Off-axis Stellar Collisions in Globular Clusters}

\begin{document}

\title{Evolution of Stellar Collision Products in Globular Clusters --
II. Off-axis Collisions.}

\author{Alison Sills\altaffilmark{1}, Joshua A. Faber\altaffilmark{2},
James C. Lombardi, Jr.\altaffilmark{3}, Frederic
A. Rasio\altaffilmark{2}, Aaron R. Warren\altaffilmark{3}}

\altaffiltext{1}{Department of Astronomy, The Ohio State University,
140 W. 18th Ave., Columbus, OH 43210;
asills@astronomy.ohio-state.edu}

\altaffiltext{2}{Department of Physics, MIT, 77 Massachusetts Ave,
Cambridge, MA 02139; jfaber@mit.edu,
rasio@mit.edu}

\altaffiltext{3}{Department of Physics and Astronomy, Vassar College,
124 Raymond Ave.,
Box 562,
Poughkeepsie, NY 12603;
lombardi@vassar.edu, aawarren00@alum.vassar.edu}

\begin{abstract}
We continue our exploration of collisionally merged stars in the blue
straggler region of the color-magnitude diagram.  We report the
results of new SPH calculations of parabolic collisions between two
main-sequence stars, with the initial structure and composition
profiles of the parent stars having been determined from stellar
evolution calculations.  Parallelization of the SPH code has permitted
much higher numerical resolution of the hydrodynamics.  We also present
evolutionary tracks for the resulting collision products, which emerge
as rapidly rotating blue stragglers.  The rotating collision products
are brighter, bluer and remain on the main sequence longer than their
non-rotating counterparts.  In addition, they retain their rapid
rotation rates throughout their main sequence
lifetime. Rotationally-induced mixing strongly affects the evolution
of the collision products, although it is not sufficient to mix the
entire star. We discuss the implications of these results for studies
of blue straggler populations in clusters. This work shows that
off-axis collision products cannot become blue stragglers unless they
lose a large fraction of their initial angular momentum. The mechanism
for this loss is not apparent, although some possibilities are
discussed.
\end{abstract}


\section{INTRODUCTION}

Globular clusters have, until recently, been studied under the
auspices of either stellar evolution or stellar dynamics.  Stellar
evolutionists often treat clusters as collections of isolated stars
all with the same age and primordial chemical abundances, while
stellar dynamicists often treat clusters as a system of point masses
moving under the influence of their mutual gravities. However, it is
becoming clear that the dynamics of a cluster can affect its
populations, and vice versa \citep{B95,PMMH99,JNR00}. Therefore, in order to
model globular clusters effectively, we need to consider both stellar
evolution and stellar dynamics, and especially their influence on each
other.

One situation of particular interest in which stellar dynamics and
stellar evolution interact involves direct stellar collisions. These
collisions, especially those involving binary stars, can have a
significant impact on the energy budget of a cluster. In particular,
they can destroy and change the properties of hard primordial
binaries. Dynamical interactions involving hard primordial binaries
are thought to be the most important source of heating that can
support a cluster against core collapse \citep[see,
e.g.,][]{GH89,Getal91,R00}. In addition, it has been suggested that,
through the indirect heating by mass loss associated with the stellar
evolution of collision products in the cluster core, collisions could
provide a mechanism for supporting a cluster against gravothermal
collapse after exhaustion of its primordial binaries \citep{GH91}. The
products of stellar collisions between main-sequence stars have the
additional useful property that they are potentially observable in the
color-magnitude diagrams of clusters as blue stragglers. Since the
population of blue stragglers is determined by the dynamics of the
cluster, we can use the blue stragglers to investigate the dynamical
state of the cluster.  For example, \cite{fer99} have suggested that
the large population of blue stragglers in M80 implies that the
cluster is close to core collapse. In order to use blue stragglers to
study globular clusters, however, we need to understand clearly the
evolution and observable properties of collision products.

In previous work \citep[hereafter Paper I]{SLBDRS97}, we used the
results of Smoothed Particle Hydrodynamics (SPH) simulations of
stellar collisions as starting models for stellar evolution
calculations, and produced detailed evolutionary models for {\it
head-on} (i.e.,\ zero impact parameter) collision products in globular
clusters. However, collisions between stars are, of course, never
exactly head-on.  \cite{LRS96} calculated collisions with non-zero
impact parameters, and some of the resulting off-axis collision
products were rotating close to their break-up speeds when they
returned to dynamical equilibrium.  The effects of rotation on the
subsequent evolution of these collision products will be
substantial. These effects have never before been included in any
theoretical study of blue straggler evolution.

In this paper, we present four new SPH simulations of off-axis
collisions, and use the results to initiate stellar evolution
calculations of the rotating collision products. In \S 2, we
outline the method and results of the SPH calculations. We summarize
our technique for combining the results of SPH simulations of stellar
collisions with detailed evolutionary models, and discuss the
modifications we have made for rotating stars. \S 3 presents the
implementation of rotation used in the Yale Rotational Evolution Code
(YREC), and the results of the evolution calculations.  We discuss the
implications of our results for blue straggler creation mechanisms and
for using blue stragglers as tracers of the dynamical evolution of
globular clusters in \S 4.

\section{HYDRODYNAMIC SIMULATIONS OF OFF-AXIS STELLAR COLLISIONS}

\subsection{Implementation of the SPH Code}

Our hydrodynamic simulations were performed using a new, parallel
version of the SPH code described in Paper I.  This code was
originally developed by \cite{R91} specifically for the study of
hydrodynamic stellar interactions such as collisions and binary
mergers \citep[see, e.g.,][]{RS91,RS92,RS94}.  SPH is a Lagrangian
hydrodynamics technique in which pressure gradient forces are
calculated by kernel estimation directly from particle positions
\citep[for a review, see][]{M92}.  In SPH simulations we solve the
equations of motion of a large number $N$ of Lagrangian fluid
particles moving under the influence of hydrodynamic forces as well as
the fluid self-gravity (in this paper $N$ is as large as
$1.05\times10^5$). For a basic overview of the SPH equations, see
\cite{RL99}.

Local densities and hydrodynamic forces at each particle position are
calculated by smoothing over $N_N$ nearest neighbors.  The size of
each particle's smoothing kernel is evolved in time to keep $N_N$
close to a predetermined optimal value.  For the high-resolution
calculations in this paper the optimal number of neighbors was set at
$N_N = 100$.  Neighbor lists for each particle are recomputed at every
iteration using a linked-list, grid-based parallel algorithm.  Note
that neighborhood is not a symmetric property, as one particle can be
included in another's neighbor list but not vice versa.  Pressure and
artificial viscosity (AV) forces are calculated by a
``gather-scatter'' method, looping over all particles in turn,
gathering the force contribution on each particle from its neighbors,
and scattering out the equal and opposite force contribution on the
neighbors from each particle.

The fluid self-gravity in our code is calculated by an FFT-based
convolution method.  The density field is placed on a 3D grid by a
cloud-in-cell method, and convolved with a kernel function calculated
once during the initialization of the simulation.  We use zero-padding
of our grids to obtain correct boundary conditions for an isolated
system, at the expense of memory storage.  Gravitational forces are
calculated from the gravitational potential by finite differencing on
the grid, and then interpolated for each particle using the same
cloud-in-cell assignment.

A number of variables are associated with each particle $i$, including
its mass $m_i$, position ${\bf r}_i$, velocity ${\bf v}_i$, entropic
variable $A_i$ and numerical smoothing length $h_i$.  The entropic
variable $A$ is a measure of the fluid's compressibility and is
closely related (but not equal) to specific entropy: for example, both
$A$ and specific entropy are conserved in the absence of shocks. For
convenience, we will refer to $A$ as entropy.  In this paper we adopt
an equation of state appropriate for a monatomic ideal gas: that is,
the adiabatic index $\gamma=5/3$ so that $P_i=A_i\rho_i^{5/3}$, where
$P_i$ and $\rho_i$ are the density and pressure of particle $i$.

The rate of increase, due to shocks, for the entropy $A_i$ of particle
$i$ is given by
\begin{equation}
{dA_i\over dt}={\gamma-1\over 2\rho_i^{\gamma-1}}\,
     \sum_jm_j\,\Pi_{ij}\,\,({\bf v}_i-{\bf v}_j)\cdot{\bf \nabla}_i
     W_{ij},
        \label{adot}
\end{equation}
where the summation is over neighbors and $W_{ij}$ is a symmetrized
smoothing kernel.  We adopt the form of AV proposed by \cite{Bal95}:
\begin{equation}
\Pi_{ij}=
\left({p_i\over\rho_i^2}+{p_j\over\rho_j^2}\right)
        \left(-\alpha \mu_{ij} + \beta \mu_{ij}^2\right),
        \label{piDB}
\end{equation}
where
\begin{equation}
\mu_{ij}=\cases{ {({\bf v}_i-{\bf v}_j)\cdot({\bf r}_i-{\bf r}_j)\over
h_{ij}(|{\bf r}_i -{\bf r}_j|^2/h_{ij}^2+\eta^2)}{f_i+f_j \over 2
c_{ij}}& if $({\bf v}_i-{\bf v}_j)\cdot({\bf r}_i-{\bf r}_j)<0$\cr
             0& if $({\bf v}_i-{\bf v}_j)\cdot({\bf r}_i-{\bf
             r}_j)\ge0$\cr}.
                \label{muDB}
\end{equation}
The terms $h_{ij}$ and $c_{ij}$ are, respectively, the average
smoothing length and sound speed associated with particles $i$ and
$j$.  Here $f_i$ is the so-called form function for particle $i$,
defined by
\begin{equation}
f_i={|{\bf \nabla}\cdot {\bf v}|_i \over |{\bf \nabla}\cdot {\bf v}|_i
+|{\bf \nabla}\times {\bf v}|_i + \eta' c_i/h_i
}, \label{fi}
\end{equation}
where
\begin{equation}
({\bf \nabla}\cdot {\bf v})_i={1 \over \rho_i}\sum_j m_j
        ({\bf v}_j-{\bf v}_i)\cdot{\bf \nabla}_i W_{ij} \label{divv}
\end{equation}
and
\begin{equation}
({\bf \nabla}\times {\bf v})_i={1 \over \rho_i}\sum_j m_j
        ({\bf v}_i-{\bf v}_j)\times{\bf \nabla}_i W_{ij}. \label{curlv}
\end{equation}
The function $f_i$ acts as a switch, approaching unity in regions of
strong compression ($|{\bf \nabla}\cdot {\bf v}|_i >>|{\bf
\nabla}\times {\bf v}|_i$) and vanishing in regions of large vorticity
($|{\bf \nabla}\times {\bf v}|_i >>|{\bf \nabla}\cdot {\bf v}|_i$).
Consequently, this AV has the advantage that it is suppressed in shear
layers.  In this paper we use $\eta^2=0.01$, $\eta'=10^{-5}$ and
$\alpha=\beta=\gamma/2$.  This choice of AV treats shocks well, while
introducing only relatively small amounts of numerical viscosity
\citep{LSRS99}.

The code was parallelized using MPI (the message passing interface),
to run efficiently on multiple processors.  Simpler methods, such as
those in HPF, prove impossible to use for SPH loops, since the
``gather-scatter'' technique, which enforces Newton's Second Law,
disallows the use of {\rm DO INDEPENDENT} loops.  In our code, the
lists of neighbors are split and divided up evenly between the
processors.  All loops over SPH particles containing sums over
quantities involving neighbors, including the AV, are distributed
among the processors, summed in part, and collected at the end of the
relevant subroutine by an {\rm MPI\_ALLGATHER} command.  The
gravitational force calculations have been parallelized using the {\it
rfftwnd\_mpi} package of the FFTW library \citep{FJ97}. All 3D grids
used by gravity routines are split up and distributed evenly among
processors in the $z$-direction, which saves a considerable amount of
computer memory.  Force contributions from grid cells contained within
each process are calculated in parallel, and gathered at the end of
the subroutine by an {\rm MPI\_ALLGATHER} command.  In benchmarking
tests, we have found that our parallel code scales very well when
using up to 32 processors on a distributed shared-memory
supercomputer.  For example, doubling the number of processors up to
32 results in an increase of total CPU time by no more than $10\%$ on
the SGI/Cray Origin2000 supercomputer.  Both the hydrodynamics and
gravitational subroutines show similar speed-up when increasing the
number of processors.

Our simulations employ equal mass SPH particles, both to keep the
resolution high in the stellar cores and to minimize spurious mixing
during the simulation; see \cite{LSRS99} for a discussion of spurious
transport induced by unequal mass particles.  Due to extreme central
densities in the parent stars, the Courant stability condition
requires exceedingly small timesteps.  Consequently, an elapsed
physical time of one hour in our high resolution calculations requires
roughly 1000 iterations and 100 CPU hours on the SGI/Cray Origin2000
supercomputer at NCSA.  The three high resolution calculations
presented in this paper each required a total of $\sim 1000$
CPU hours.

\subsection{Initial Conditions}

\begin{figure}[ht!]
\plotone{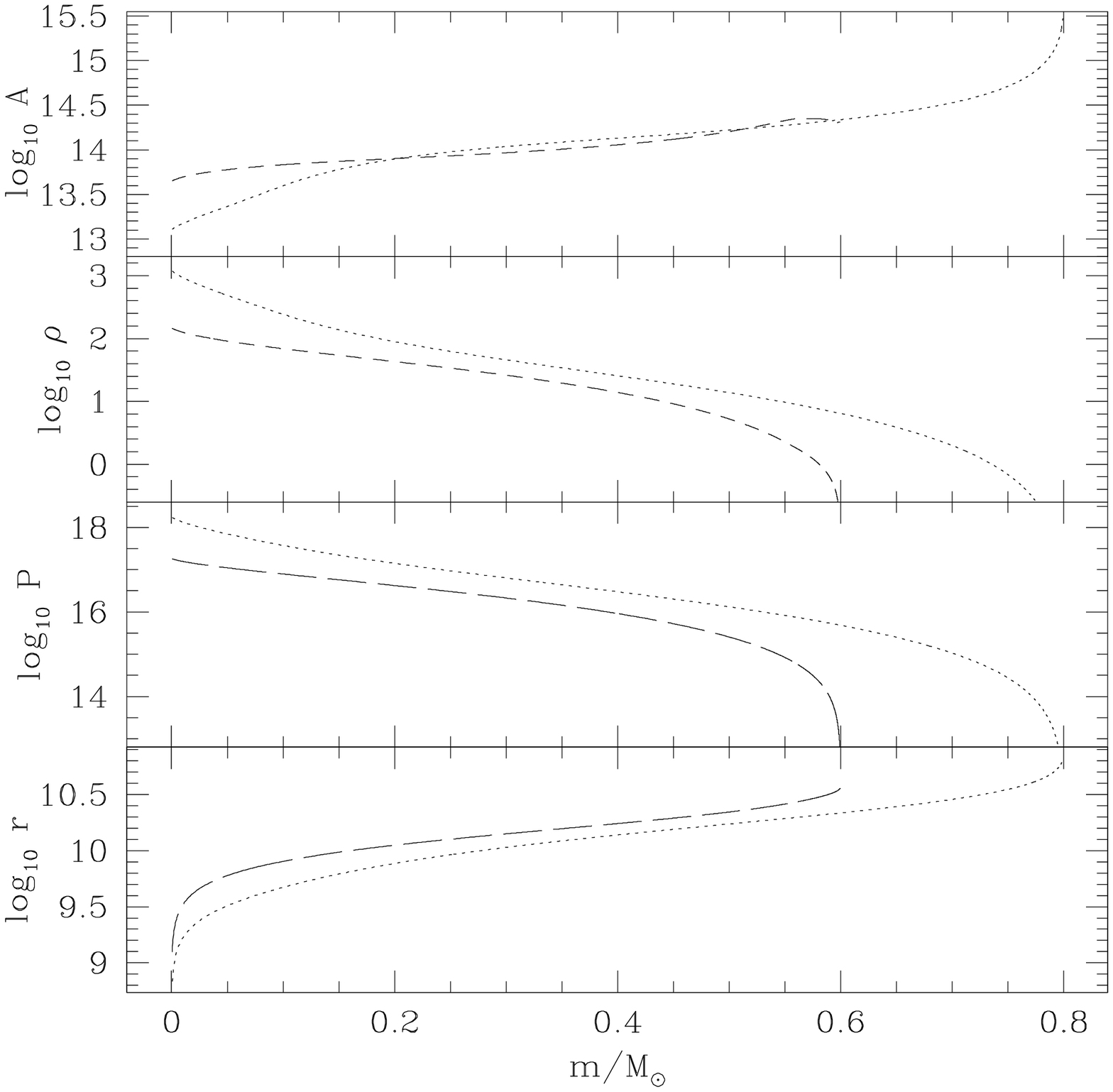}
\caption {Profiles of the entropy $A$, pressure $P$, density $\rho$
and radius $r$ as a function of enclosed mass $m$ for realistically
modeled $0.6M_\odot$ (dashed curve) and $0.8M_\odot$ (dotted curve)
parent stars. Units are cgs.
\label{logthermpar}
}
\end{figure}

All of our hydrodynamic simulations use parent star models whose
characteristics are based on the results of YREC calculations, as
discussed in \cite{SL97}. In particular, we evolved (non-rotating)
main-sequence stars of total mass $M=0.6$ and $0.8 M_\odot$ with a
primordial helium abundance $Y=0.25$ and metallicity $Z=0.001$ for 15
Gyr, the amount of time needed to exhaust the hydrogen in the center
of the $0.8M_\odot$ star.  The thermodynamic and structural profiles
of the parent stars are shown in Figure \ref{logthermpar}.  The total
helium mass fractions for the $0.6$ and $0.8 M_\odot$ parent stars are
0.286 and 0.395, and their radii are $0.52 R_\odot$ and $0.96
R_\odot$, respectively.  From the pressure and density profiles of
these models, we compute the entropy profile and assign values of $A$
to SPH particles accordingly.  In addition, the chemical abundance
profiles of 15 different elements are used to set the composition of
the SPH particles.  To minimize numerical noise, each parent star's
SPH model is relaxed to equilibrium using an artificial drag force on
the particles, and then these relaxed models are used to initiate the
collision calculations.

The stars are initially non-rotating and separated by 5 $R_{TO}$,
where $R_{TO}=0.96R_\odot$ is the radius of a turnoff star.  The
initial velocities are calculated by approximating the stars as point
masses on an orbit with zero orbital energy (i.e., we assume the
parabolic limit, appropriate for globular clusters) and a pericenter
separation $r_p$.  A Cartesian coordinate system is chosen such that
these hypothetical point masses of mass $M_1$ and $M_2$ would reach
pericenter at positions $x_i=(-1)^{i}(1-M_i/(M_1+M_2))r_p$,
$y_i=z_i=0$, where $i=1,2$ and $i=1$ refers to the more massive star.
The orbital plane is chosen to be $z=0$.  With these choices, the
center of mass resides at the the origin.

\subsection{Hydrodynamic Results and Conversion to YREC Format
\label{hydrodynamicResults}}

Table \ref{tbl-summary} summarizes the parameters and results of all
of our SPH simulations.  The first column gives the name by which the
calculation is referred to in this paper; we use lower case letters to
distinguish the present calculations from the corresponding
calculations involving polytropic parent stars presented in
\cite{LRS96}.  The second and third columns give the masses $M_1$ and
$M_2$ of the parent stars. Column~(4) gives the ratio $r_p/(R_1
+R_2)$, where $r_p$ is the pericenter separation for the initial orbit
and $R_1+R_2$ is the sum of the two (unperturbed) stellar radii. This
ratio has the value $0$ for a head-on collision, and $\sim 1$ for a
grazing encounter.  Column~(5) gives the number of SPH particles.
Column~(6) gives the final time $t_f$ at which the calculation was
terminated.  Column~(7) gives the total angular momentum of the merger
remnant.  Column~(8) gives the ratio $T/|W|$ of rotational kinetic
energy to gravitational binding energy of the (bound) merger remnant
in its center-of-mass frame at time $t_f$.  Column~(9) gives the mass
of the merger remnant (the gravitationally bound fluid).  Columns~(10)
and~(11) give the velocity components $V_x$ and $V_y$ for the merger
remnant's center of mass at time $t_f$ in the system's center-of-mass
frame.  Since the amount of mass ejected during a parabolic collision
is small, the merger remnant never acquires a large recoil
velocity.  The case e, f and k simulations implemented $N_N=100$
neighbors
per particle, while case j$^\prime$ implemented $N_N=32$ neighbors.
Case j$^\prime$ is a low resolution
calculation of a nearly head-on collision.

Figure \ref{color} illustrates the dynamical evolution for case~e: a
turnoff main-sequence star ($M_1=0.8M_\odot$) collides
with a slightly less massive star
($M_2=0.6M_\odot$). The parabolic trajectory has a
pericenter separation $r_p=0.25 (R_1+R_2)$.  The first collision at
time $t\simeq2$ hours disrupts the outer layers of the two stars, but
leaves their inner cores essentially undisturbed.  The two components
withdraw to apocenter at $t\simeq 3.2$ hours before colliding for the
second, and final, time.  The merger remnant is rapidly and
differentially rotating, and shear makes the bound fluid quickly
approach axisymmetry within a few dynamical timescales.

\begin{figure}[ht!]
\plotone{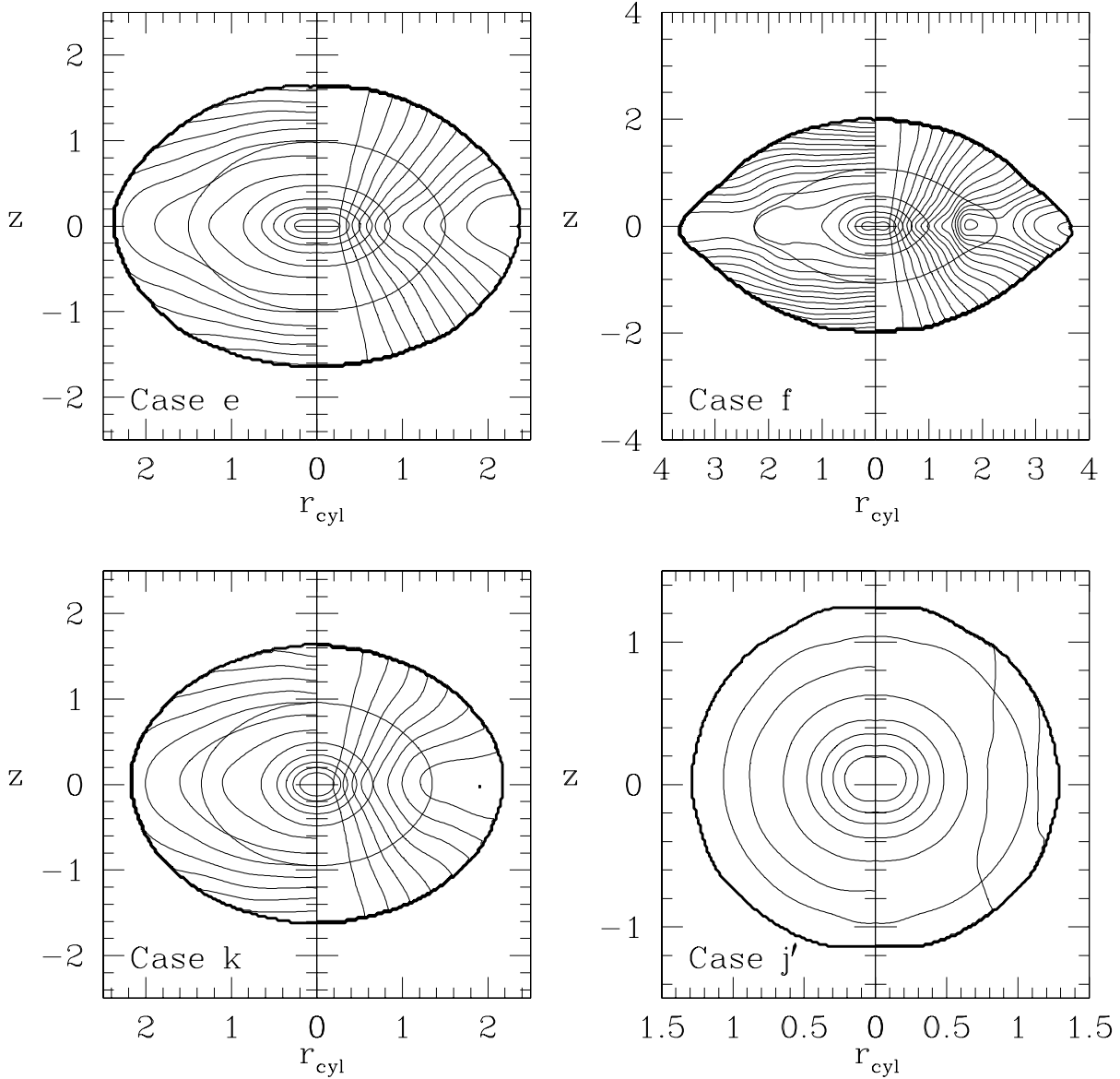}
\caption {Constant density, specific angular momentum and entropy
contours at the end of each simulation in slices containing the
remnants' rotation axes.  Here $r_{cyl}$ is the cylindrical radius
measured from the rotation axis, in units of the radius of a turnoff
mass star. The closed loops that extend to both the right and left
halves of each plot correspond to the isodensity surfaces enclosing
15\%, 30\%, 45\%, 60\%, 75\% and 90\% of the remnant mass; the thick
outermost bounding curve correspond to an enclosed mass fraction
$m/M=0.95$.  The left half of each plot shows constant entropy
contours, with a linear spacing of $1.04\times 10^{15}$ (cgs), while
the right half of each plot shows constant specific angular momentum
contours, with a linear spacing of $2.66\times 10^{17}$ cm$^2$
s$^{-1}$.
\label{all}
}
\end{figure}

Figure \ref{all} shows contours of the density $\rho$, the entropy
$A$, and the $z-$component of the specific angular momentum $j=\omega
r_{cyl}^2$ for each collision product in a slice containing its
rotation axis.  Here $r_{cyl}$ is the cylindrical radius measured from
the rotation axis.  The contours shown represent an average over the
azimuthal angle, as the product approaches axisymmetry at the end of
the simulation.  By definition, the collision products are not
barotropes since the specific angular momentum $j$ (and hence
$\omega$) has a distinct dependence on $z$.  For a rotating star,
stable thermal equilibrium requires $d\omega/dz=0$ in chemically
homogeneous regions, as discussed in \S \ref{evolutionaryResults}.
From the set of specific angular momentum contours presented in Figure
\ref{all}, it is therefore evident that the outer layers of the
collision products (where the composition gradient is small) are not
in thermal equilibrium.  The shape of the specific angular momentum
contours is due to the tendency of dynamical shear to equalize the
angular velocity $\omega$ on surfaces of constant density
\citep{PKSD89}, so that $j$ increases with $z$ at fixed $r_{cyl}$.

To convert our three-dimensional hydrodynamic results into the
one-dimensional YREC format, we begin by averaging the entropy and
specific angular momentum values in $\sim25$ bins in enclosed mass
fraction.  The resulting profiles are extrapolated over the outermost
$\sim$5\% of the remnant mass, since in practice the simulations must
be terminated before all bound fluid has fallen back to the remnant's
surface \citep[for a discussion of how the values of the final time
$t_f$ are chosen, see \S 2.3 of][]{LRS96}.  The specific angular
momentum profile is then renormalized (by a factor of close to unity),
to ensure that the extrapolation does not modify the total angular
momentum of the remnant.  With the $A$ and $j$ profiles given, the
structure of the rotating remnant is uniquely determined in the
formalism of \cite{ES76}, by integrating the general form of the
equation of hydrostatic equilibrium [see their eq. (9)].  To do so, we
implement an iterative procedure in which initial guesses at the
central pressure and angular velocity are refined until a
self-consistent YREC model is converged upon.  The temperature profile
is calculated from the ideal gas equation of state, while the
luminosity profile is calculated according to the equation of
radiative transport.  This first approximation for the structure of
the star was used as a starting model for the stellar evolution code.

\begin{figure}[ht!]
\plotone{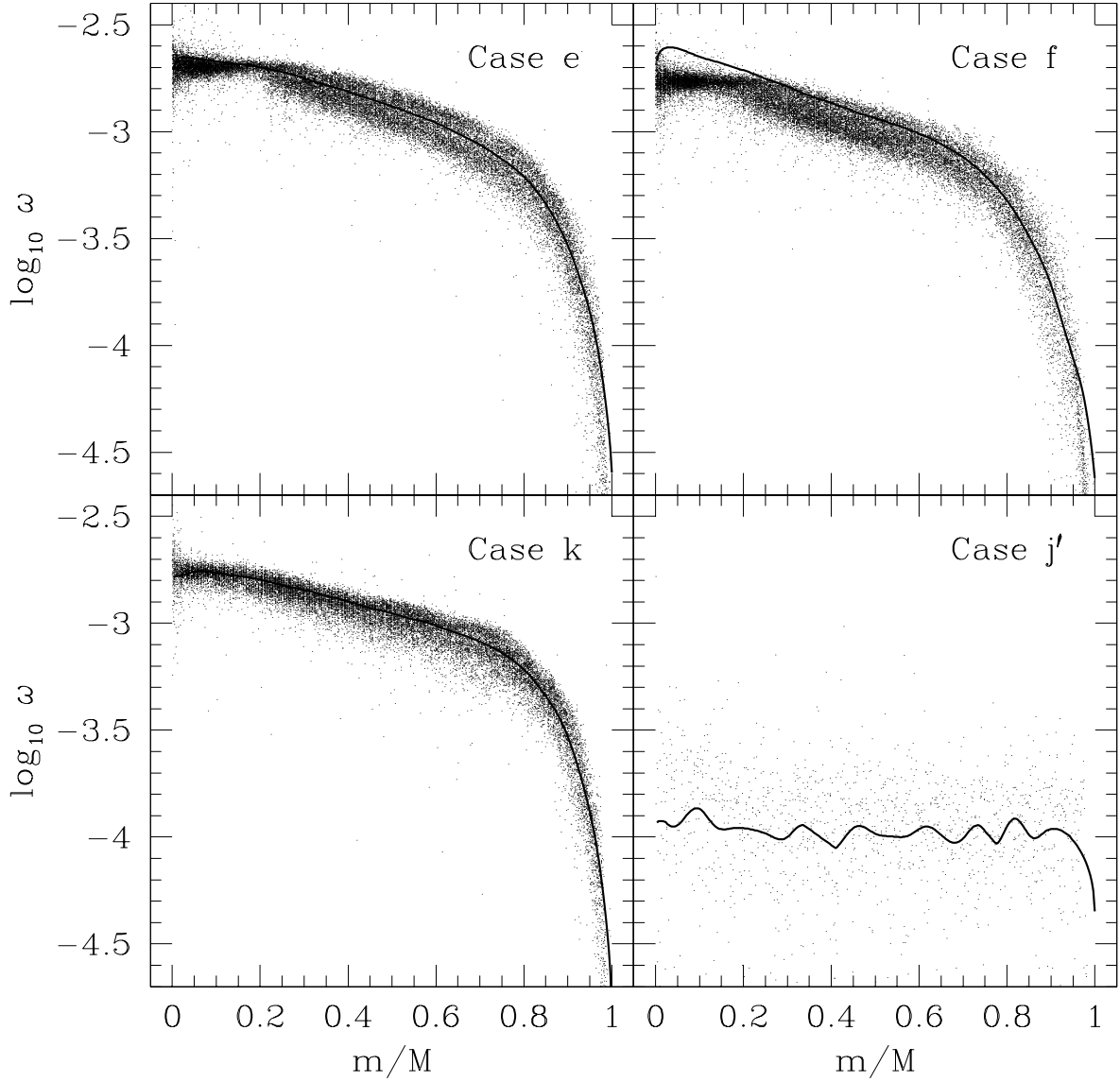}
\caption { Angular velocity $\omega$ as a function of the enclosed
mass fraction $m/M$ in the final merger remnants.  The points
represent particle values from an SPH simulation, with only half of
the particles displayed in cases e, f and k.  The solid lines show the
$\omega$ profiles of the initial YREC models, generated by a
procedure that constrains the entropy and specific angular momentum
profiles of each model to be that given by the SPH results.  Units of
$\omega$ are rad$\,{\rm s}^{-1}$.
\label{alllogomegam}
}
\end{figure}

Although the pressure, density, and angular velocity profiles of the
resulting YREC model cannot also be simultaneously constrained to
equal exactly the corresponding (averaged) SPH profiles, the
differences are slight (see, for example, the $\omega$ profiles in
Figure \ref{alllogomegam}).  Indeed, the subsequent evolutionary
tracks are not significantly different for remnant models that use the
form of the $\omega$ profile taken directly from SPH.  The decision to
constrain the $j$ profile, as opposed to the angular velocity $\omega$
profile, guarantees that the total angular momentum of the initial
YREC model agrees with that from the SPH results.  Furthermore, the
SPH $\omega$ profile can be slightly less accurate than the $j$
profile, since the latter is less sensitive to the resolution of the
calculation.

\section{EVOLUTION OF THE ROTATING COLLISION PRODUCTS}

\subsection{Implementation of the Stellar Evolution Code}

Our stellar evolution calculations are performed with YREC. YREC is a
one-dimensional code, in which the star is divided into shells along
surfaces of constant gravitational plus rotational potential. The code
solves the equations of stellar structure with the Henyey technique,
and follows the rotational evolution with the formalism of
\cite{ES76}.  \cite{GDKP92} give a detailed description of the
physics implemented in the evolution code. We used the same opacities,
equation of state and model atmospheres as in Paper I. The free
parameters in the code (the mixing length and parameters that set the
efficiency of angular momentum transport and rotational chemical
mixing) are set by calibrating a solar mass and solar metallicity
model to the sun.

Rotation is treated by evaluating physical quantities on equipotential
surfaces rather than the spherical surfaces usually used in stellar
models. The hydrostatic equilibrium and radiative transport equations
contain terms that account for the lack of spherical symmetry in the
rotating star. A number of rotational instabilities that transport
angular momentum and material within the star are followed, including
dynamical shear \citep{PKSD89}, meridional circulation \citep{vZ24},
secular shear \citep{Z74}, and the Goldreich-Schubert-Fricke (GSF)
instability \citep{GS67,F68}. Angular momentum transport and the
associated chemical mixing are treated as diffusion processes, with
diffusion coefficients that account for each active mechanism within
unstable regions of the star.  The diffusion coefficients are
proportional to circulation velocities, which have been estimated by
\cite{ES78}.  In addition to the internal rearrangement of angular
momentum, angular momentum also can be drained from an outer
convection zone through a magnetic wind, using the formalism given by
\cite{CDP95}. The Endal \& Sofia scheme is valid across a wide range
of rotation rates, for a restricted class of angular momentum
distributions. This scheme requires that the angular velocity is
constant on equipotential surfaces, which does not allow for modeling
of latitude-dependent angular velocity profiles.  See \cite{MM97} for
a detailed discussion of the validity of this approach.  For a
detailed description of the implementation of rotation in YREC, see
\cite{PKSD89}.

\subsection{Evolutionary Results \label{evolutionaryResults}}

Although in dynamical equilibrium, the nascent collision products are
initially far from thermal equilibrium and contract on a thermal
timescale.  During this stage, their luminosity is predominately
provided by the conversion of gravitational potential energy,
analogous to the situation in pre-main-sequence stars. After the
contraction, central hydrogen burning quickly becomes significant, and
we refer to this point as the zero age main sequence (ZAMS). Note,
however, that unlike normal ZAMS stars, these stars are far from being
chemically homogeneous.  Many qualitative features of the subsequent
evolution are similar to that of normal intermediate mass stars; for
example, it is possible to define the subgiant and giant stages of
evolution.  Nevertheless, the timescales for this evolution, and even
the actual evolutionary tracks, can be substantially different from
those of normal stars, because of the initial structure of the
collision product and because of the effects of the high rotation
rate.

The initial total angular momentum of the collision products (Table
\ref{tbl-summary}) is as much as 10 times larger than that of normal
stars of comparable mass at the pre-main-sequence birth line. The
stellar collision products differ substantially from normal
pre-main-sequence stars in another significant way. Normal pre-main-sequence stars have large convective envelopes, and therefore can lose
substantial amounts of angular momentum via a magnetic wind. The sun
has lost about 99\% of its initial angular momentum, for example
\citep{PKSD89}. However, the stellar collision products are not
convective at the end of the SPH simulation, nor do they develop
convection zones during the thermal readjustment and main-sequence
stages.  Therefore, the collision products do not lose any angular
momentum by a magnetic wind until the subgiant branch.

Rotational instabilities transport angular momentum from the center of
the star to the surface approximately on an Eddington-Sweet timescale
(the thermal timescale divided by the ratio of centrifugal
acceleration to gravitational acceleration).  For the rapidly rotating
stellar collision products, this timescale is on the order of only
$10^7$ years for the stars as a whole, and is even shorter in the
outer layers.  The dominant mixing mechanism in these stars is the GSF
instability, which addresses axisymmetric perturbations: A structure
in which the angular velocity $\omega$ is constant on cylinders is the
only thermally stable configuration in regions of homogeneous chemical
composition. Since dynamical shear acts to remove any latitude
dependence of the angular velocity \citep[see][]{PKSD89}, regions that
are subject to the GSF instability evolve towards rigid rotation. In
the case of our stellar collision products, the angular velocity
profile is initially a steeply decreasing function of enclosed mass
fraction (Figure \ref{alllogomegam}).  Therefore, the GSF instability
serves to raise the surface rotation rate and decrease the interior
rotation rate, flattening the $\omega$ profile. Since the timescale
for this to occur is shortest at the surface, the $\omega$ profile
begins to flatten there, and then the flattened profile penetrates
inwards. The star has a very large reservoir of angular momentum, and
this redistribution of angular momentum can cause the surface of the
star to rotate faster than its breakup velocity.

An obvious solution here would be simply to remove the mass from the
calculation as it becomes unbound to the star. Unfortunately, under
this scheme, rapidly rotating stellar collision products could be
completely disrupted. The total angular momentum of the star is so
large that if the star were rotating as a solid body, its rotational
velocity would be more than its breakup velocity. This is true
immediately following the collision, when the star has its largest
radius, and is even more obvious as the star collapses to the main
sequence. The breakup velocity is proportional to $R^{-1/2}$, so it
increases as the radius $R$ decreases. However, the surface rotation
rate is proportional to $R^{-1}$, and increases faster than the
breakup velocity as $R$ decreases.  As the GSF instability works to
flatten the rotation profile further and further into the center of
the star, more of the star becomes unbound, and eventually the entire
star can be spun apart.  This poses a problem for the theory that blue
stragglers are created via stellar collisions: either blue stragglers
are not created through physical off-axis stellar collisions, or some
mechanism(s) can remove angular momentum from the star on short
timescales. Possible mechanisms will be discussed in \S 4.

\begin{figure}
\plotone{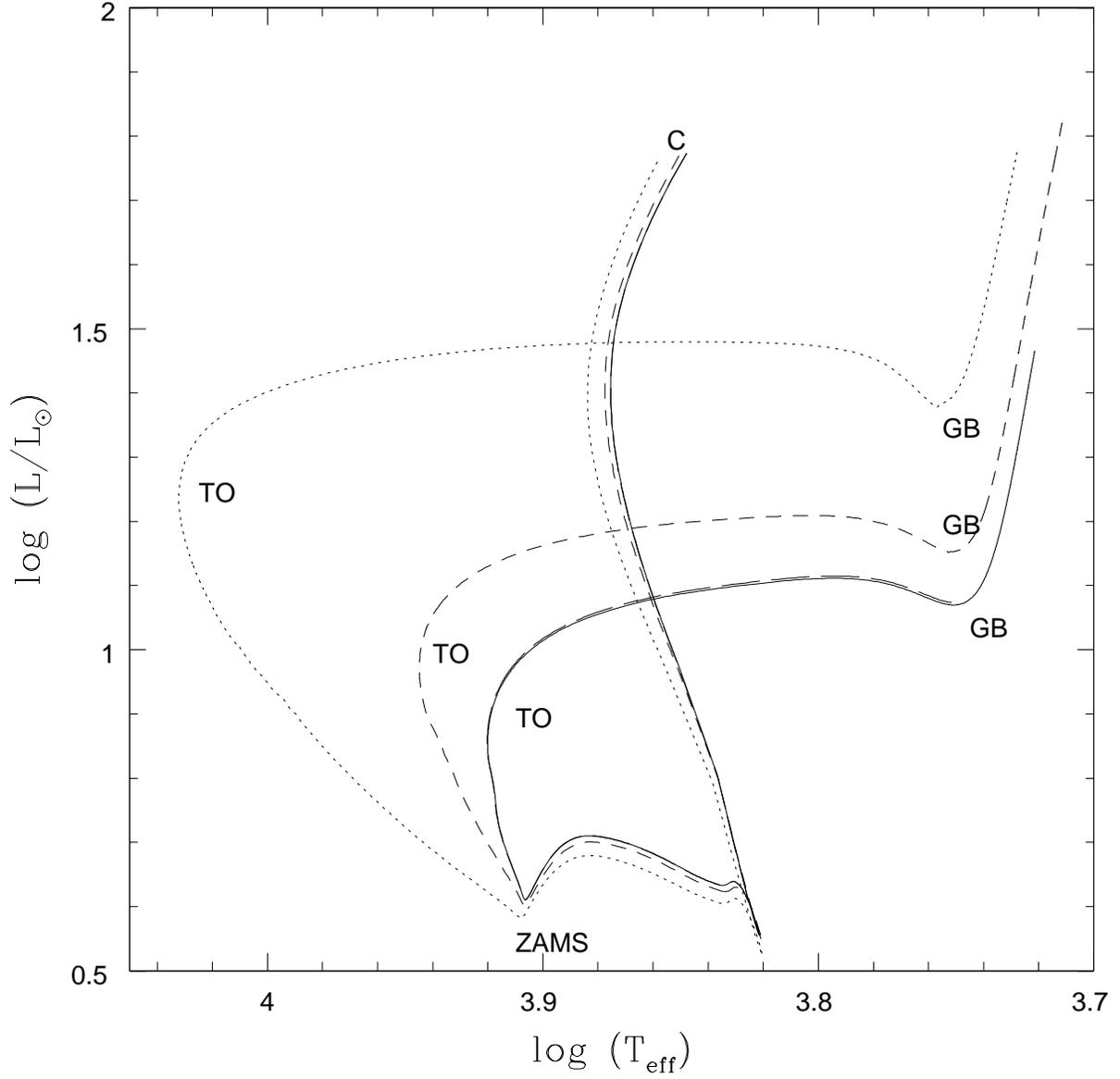}
\caption {Evolutionary tracks for case k with the initial angular
velocity $\omega_0$ divided by 5 (dotted line), 10 (short dashed
line), 100 (long dashed line) and 1000 (solid line line). The end of
the collision is marked with a 'C', the start of central hydrogen
burning with 'ZAMS', the turnoff with 'TO', and the base of the giant
branch with 'GB'. \label{caseKdiv}}
\end{figure}

For now we will simply postulate that the angular momentum of the
collision products can be reduced. In order to study the effects of
rotation on stellar collision products, we performed some experiments
on case k ($0.6 M_{\sun} + 0.6 M_{\sun}, r_p=0.25 (R_1+R_2)$).  We
artificially decreased the initial rotation velocity of each shell by
a large factor before evolving the model.  In particular, we divided
the rotation velocities by 1000, 100, 10, 5 and 2, which reduced their
total angular momentum by almost the same factors. In this last case,
the star still had enough angular momentum that it began to rotate
faster than its breakup velocity. The evolutionary tracks for the
other angular momentum values are shown in Figure \ref{caseKdiv}. The
tracks with the least angular momentum are not noticeably different
from the evolutionary track of the corresponding non-rotating model.
As the total angular momentum of the star increases, the star spends
more time on the main sequence due to two effects. First, rotation
provides an additional source of support for the star, so that it does
not have to produce as much pressure through nuclear burning to
prevent its collapse. Second, rotationally induced mixing moves
hydrogen into the core of the star, increasing the amount of available
fuel.

\begin{figure}
\plotone{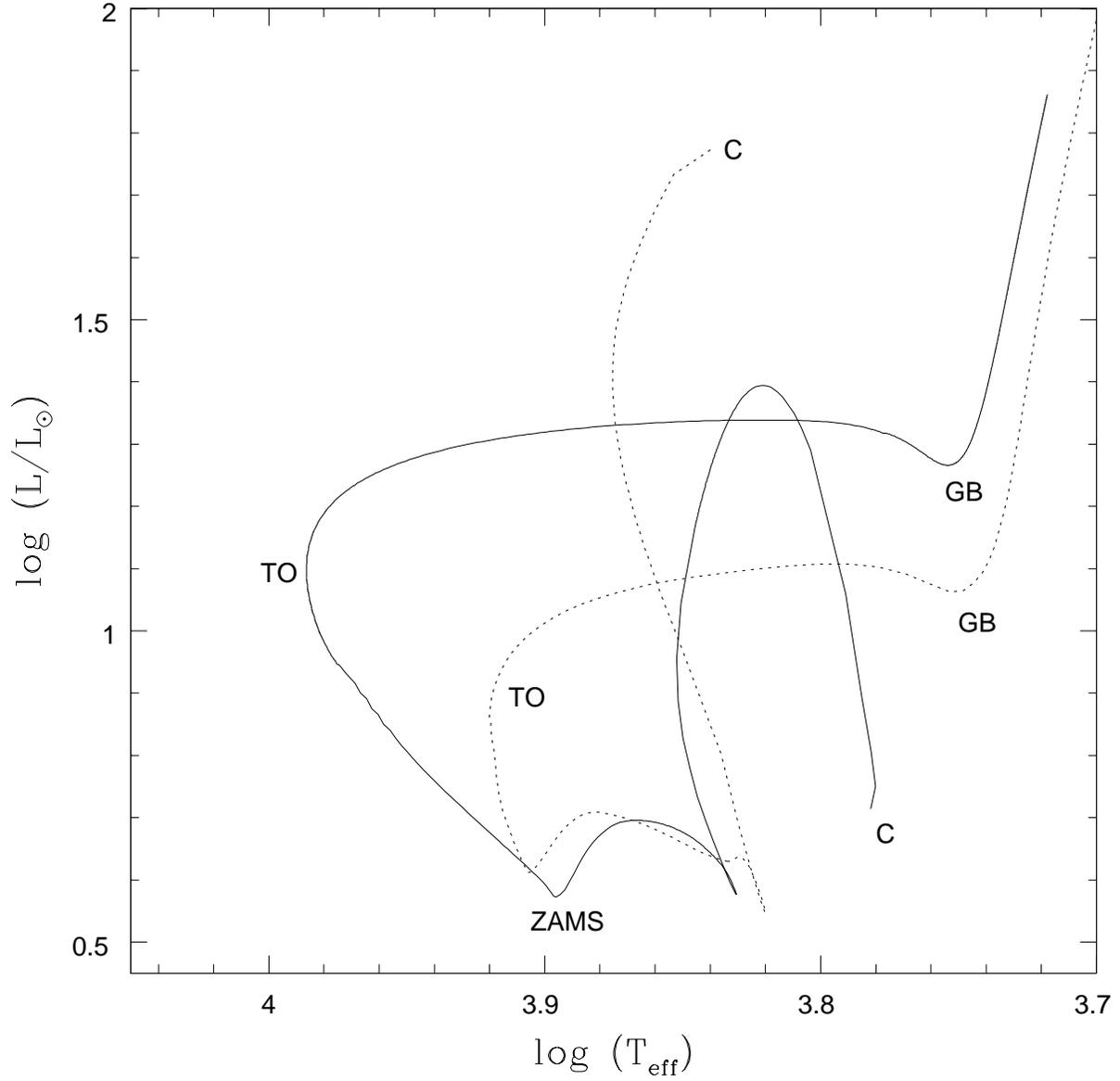}
\caption {Evolutionary track for case j$^\prime$ (solid line), and for
the same case with the same initial structure profiles, but zero
angular momentum (dotted line). The different evolutionary points are
marked as in Figure \ref{caseKdiv}. \label{jprime}}
\end{figure}

We also evolved the product of the low resolution SPH simulation of a
barely off-axis collision (case j$^\prime$). In this situation, two
0.6 $M_{\sun}$ parent stars collide with an impact parameter of only
$r_p = 0.02 R$, where $R=0.52 R_\odot$. The SPH results are used
directly as the initial conditions for the evolution code as described
in \S \ref{hydrodynamicResults}, without rescaling the angular
velocity profile.  This collision has sufficiently little angular
momentum that even if the collision product rotated as a solid body,
it would rotate slower than its breakup velocity.  We used this case
to confirm that the results of SPH simulations could indeed be evolved
directly, and to see what the effect of even a barely off-axis
collision was on the subsequent evolution of the collision
product. The results are shown in Figure \ref{jprime}. For comparison,
a non-rotating version of the same star is also shown. The
non-rotating star was evolved using the same starting model as the
rotating case, but with zero angular momentum. The effects of
rotation, even for a collision that is barely off-axis, are
remarkable. This collision product has an initial total angular
momentum of $2.1 \times 10^{50}$ g cm$^2$ s$^{-1}$, comparable to that
of a pre-main-sequence star of the same mass. Since the collision
product does not lose angular momentum during its evolution through
the main-sequence stage, its rotational velocity remains quite large
(272 km s$^{-1}$ at the turnoff). Substantial rotational mixing
occurs, prolonging its main-sequence lifetime to 3.7 Gyr (compared to
1.8 Gyr for the equivalent non-rotating star).

The evolutionary tracks for the case e, f and k collision products are
shown in Figure \ref{evolution}. Here we have reduced the initial
angular velocities by a factor of 5, in order to prevent the stars
from rotating faster than breakup. None of the other profiles are
changed initially, although YREC converges on a solution which
satisfies the rotational equations of stellar structure.  Some details
of the various evolutionary tracks are given in Table
\ref{tbl-evolution}.  Cases e and f involve the same parent stars (0.6
$M_\odot$ and 0.8 $M_\odot$) but have different pericenter
separations, namely 0.25 and 0.5 $(R_1 + R_2)$, respectively.  Since
case f has significantly more angular momentum than case e, much more
mixing occurs, and the evolutionary tracks are substantially
different.  The track of the case f collision product also shows the
most striking differences from the track of a normal, slowly rotating
star.  Although not even the case f collision product is fully mixed,
hydrogen is carried into the core nearly as fast as it is depleted by
nuclear burning, extending the main-sequence lifetime in case f by a
factor of 5 over that of case e.

\begin{figure}
\plotone{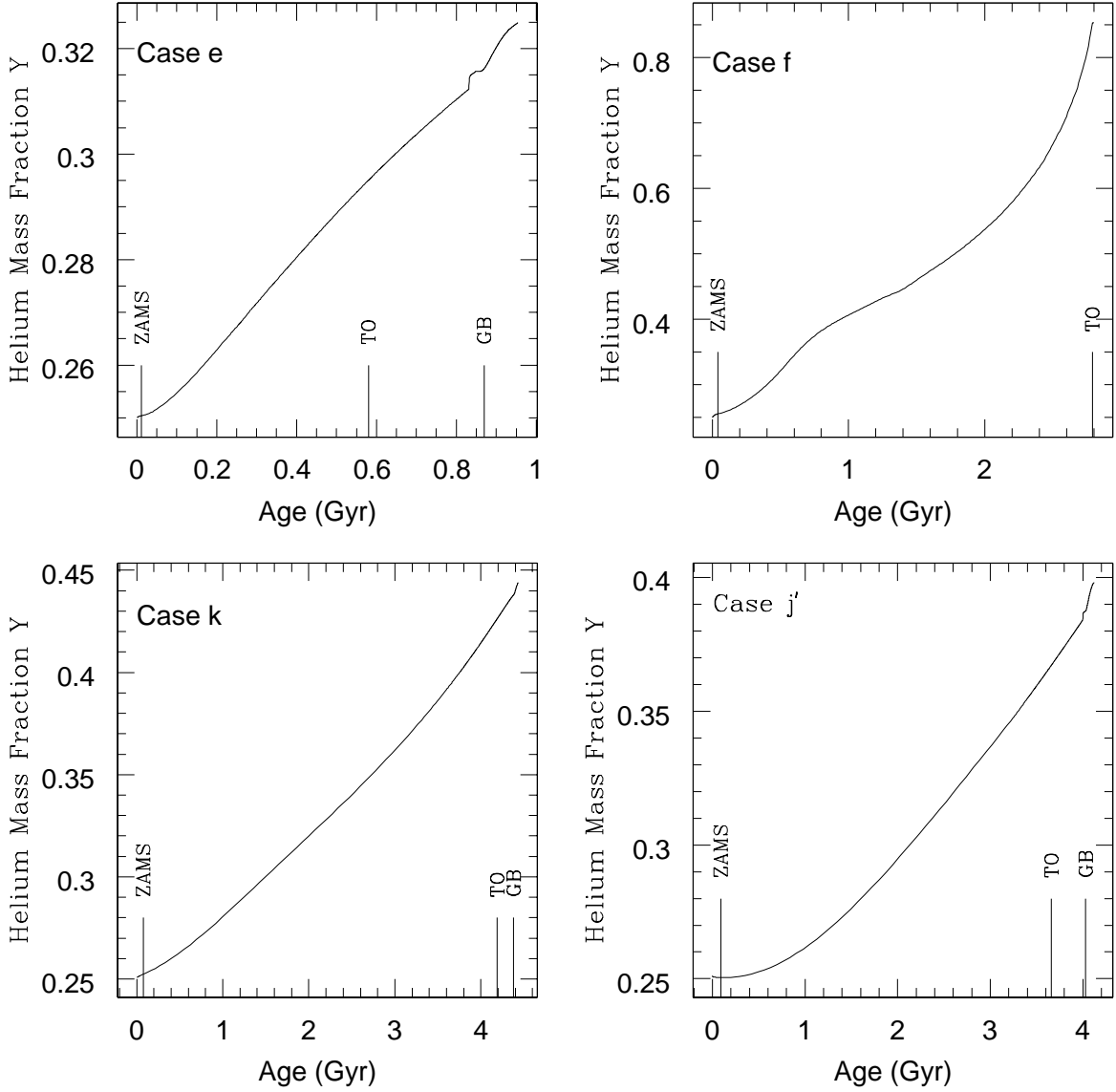}
\caption {Surface helium abundance as a function of age for each of
the four cases: cases e, f and k with the initial rotational velocity
divided by 5 in each case, and case j$^\prime$. The ages of the main
sequence, the turnoff and the base of the giant branch are
marked. \label{surfHe}}
\end{figure}

In addition to transporting angular momentum and carrying hydrogen
into the core, rotational instabilities also carry helium to the
surface, affecting the surface opacity and causing the star to become
continuously brighter and bluer while on the main sequence.  The most
dramatic example of this effect can be seem in the case f track of
Figure \ref{evolution}, which reaches a luminosity of $300 L_\odot$
and a surface temperature of $2.5\times 10^4$ K while still on the
main sequence.  In this case, the helium mass fraction at the surface
reaches an incredible $Y=0.85$.  Figure \ref{surfHe} shows the surface
abundance of helium as a function of time for the collision products
featured in this paper.

The differences between the case e and case f tracks, which correspond
to collision products of nearly the same mass, help to emphasize that
the total mass of a collision product is not the primary factor in
determining its relative position on a Hertzsprung-Russell (HR)
diagram, unlike the situation for normal main-sequence stars.  As a
second example, consider the collision product of case k, which has a
mass of only $1.16 M_\odot$, but which is nevertheless brighter and
bluer at turnoff than the $1.36 M_\odot$ product of case e.  As
discussed in Paper I and \cite{SL97}, a collision that involves a
turnoff mass star (such as in cases e and f) will produce a blue
straggler with a hydrogen-depleted core, since the core of the turnoff
star sinks to the center of the collision product.  Therefore, the
main-sequence lifetime for the higher mass products is generally
shorter.  In lower mass collisional blue stragglers, there is then
more time for helium to be mixed to the surface, and the surface
helium abundance can grow larger than in the higher mass collision
products (compare cases e and k in Figure \ref{surfHe}).  As a result,
a lower mass collision product can become brighter and bluer than one
of higher mass.  Consequently, blue straggler masses that have been
determined from color-magnitude diagram position could be in error if
the blue stragglers are rapidly rotating.

\begin{figure}
\plotone{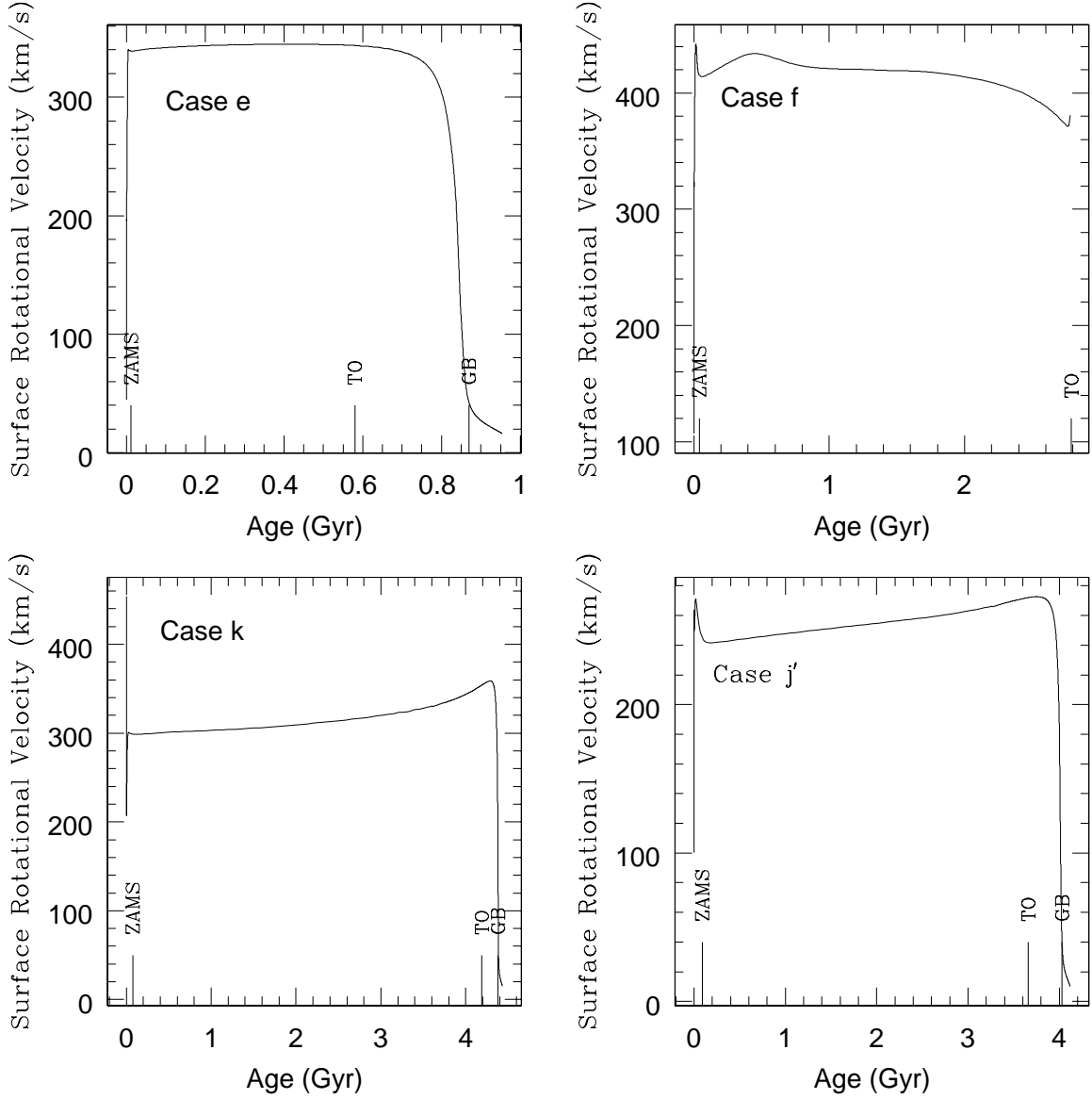}
\caption {Surface rotational velocity as a function of time for four
cases: case e with $\omega_0/5$, case f with $\omega_0/5$,
case k with $\omega_0/5$, and case j$^\prime$. \label{vrottime}}
\end{figure}

Figure \ref{vrottime} shows the surface rotational velocity as a
function of time for the four cases featured in this paper. The ages
of the ZAMS, the turnoff, and the base of the giant branch are marked
in each of the panels. All four cases show similar behavior: the
rotational velocity adjusts rapidly as the star contracts to the main
sequence. Once on the main sequence, the surface rotational velocity
remains approximately constant. As the star becomes a subgiant, it
develops a surface convection zone, and the standard angular momentum
loss mechanism (via a magnetic wind) begins to operate. The loss rate
is proportional to the surface angular velocity to the third power, so
these rapidly rotating stars very efficiently shed most of their
angular momentum before reaching the giant branch.

\section{SUMMARY AND DISCUSSION}

We have modeled blue stragglers created in off-axis collisions between
cluster main-sequence stars. Since even a minutely off-axis collision
can produce a rapidly rotating star, most collision products cannot
remain gravitationally bound as they contract to the main sequence
unless they somehow lose angular momentum. The standard angular
momentum loss mechanism (via a magnetic wind) does not operate in
collisional blue stragglers, since magnetic braking is effective only
in stars with surface convection zones deep enough to support a
dynamo. The stellar collision products lack a surface convection zone,
even during their collapse to the main sequence. Therefore, either
blue stragglers cannot be created by stellar collisions, or some other
mechanism for removing angular momentum must occur.

Angular momentum can be lost from stars in a variety of ways.  Stars
more massive than these collision products have substantial mass loss
rates during their main-sequence evolution, which can be on the order
of $10^{-9}M_{\sun}$ yr$^{-1}$ for stars of 3 $M_{\sun}$ \citep{L81}.
Such a high mass loss rate will remove a substantial amount of angular
momentum from a star. The mass loss in these stars takes the form of a
line radiation-driven wind \citep{CAK75}, which requires high
luminosities in order to remove a large amount of mass.  This
mechanism may be of importance in extremely bright collision products
(such as in case f), although low luminosity blue stragglers are
unlikely to be significantly affected.

If the collision product is a member of a binary system, it can lose
angular momentum through the orbital evolution of the system. Angular
momentum loss is usually discussed in the context of contact or
near-contact binary systems, and requires that one of the stars has a
stellar wind and a co-rotating magnetic field \citep{S95}. It is
possible to have a stellar collision in which the product has a binary
companion (through an encounter of a single star with a binary system
\citep{D95}, or two binary systems with each other
\citep{BSD96}. However, it is not clear that the resulting binary
system will be in contact, and it is highly unlikely that all blue
stragglers formed in such a fashion.

The most plausible mechanism for removing angular momentum from these
stellar collision products is disk-locking. In low mass pre-main-sequence stars, the torque placed on a star through a magnetic field
coupled to a disk of material at a few stellar radii is enough to slow
the rotation of these stars by as much as an order of magnitude
\citep{SPT00}.  This mechanism requires the presence of both a
magnetic field and a disk of material around the star.  The SPH
simulations of the stellar collisions show that typically a few
percent of the mass of the system is not bound to the final collision
product. However, this mass will leave the system on a dynamical time
(on the order of hours) and cannot become the disk. A more reasonable
source of disk mass is the outer layers of the star that are spun off
during the initial contraction on a thermal timescale. Since the
amount of mass necessary for such a disk is on the order of a few
hundredths of a solar mass, the disk-locking could begin very soon
after the collision, and remove the angular momentum quite quickly. A
simple calculation shows that this scenario is not unreasonable. We
assume that angular momentum is carried away by mass which is coupled
to the surface rotation rate of the star down to the Alfven radius
\citep{K91}. In order to lose 80\% of the star's angular momentum (in
case f), we need to lose only $\sim0.1 M_{\sun}$.  The other
ingredient in this scenario is the presence of a magnetic field.  We
know that low mass stars have substantial magnetic fields. It is not
clear what happens to those fields after the stellar collision, and
whether they will be coherent enough to lock the star to a disk.  More
work is necessary to determine if this or any other angular momentum
loss mechanism occurs in stellar collision products.

We assumed that the stellar collision products would lose a large
fraction (up to $\sim$80\%) of their initial angular momentum early in
their evolution. We then continued their evolution through the main
sequence to the base of the giant branch. The rotating stars are bluer
and brighter than their non-rotating counterparts, and have longer
main-sequence lifetimes. Both of these changes to the evolutionary
tracks will have an effect on the derived blue straggler distributions
in the color-magnitude diagram, and therefore will change our
conclusions about the dynamical state of globular clusters.

Since rotating blue stragglers live on the main sequence longer, we
expect that their distributions will be more populated near the main
sequence than those presented in \cite{SB99}.  Since the observed
distributions of blue stragglers in the color-magnitude diagram are
already extended to cooler temperatures than the predicted
distributions, the increase in blue straggler main-sequence lifetime
will exacerbate the lack of agreement between observations and theory.
It is possible that the width of the observed distribution is a
selection effect, caused by the blended light of chance superpositions
or true binary stars. More careful observations of blue straggler
populations in globular clusters may clear up this particular mystery.

While none of the stars investigated in this work are fully mixed,
rotation does cause helium to be mixed to the surface. Rotating blue
stragglers tend to be
much brighter than their non-rotating counterparts, and much
hotter. Therefore, rotation could be responsible for anomalously
bright blue stragglers, such as the central 6 blue stragglers in
NGC 6397 \citep{Lau92} or the brightest three blue stragglers in M3
\citep{fer97}. Detailed comparisons between observed distributions and
theoretical distributions based on rotational tracks are necessary to
determine if there are two distinct populations of collisional blue
stragglers in globular clusters, or if all blue stragglers can be
explained by a population with a range of collisional impact
parameters.

\begin{figure}
\plotone{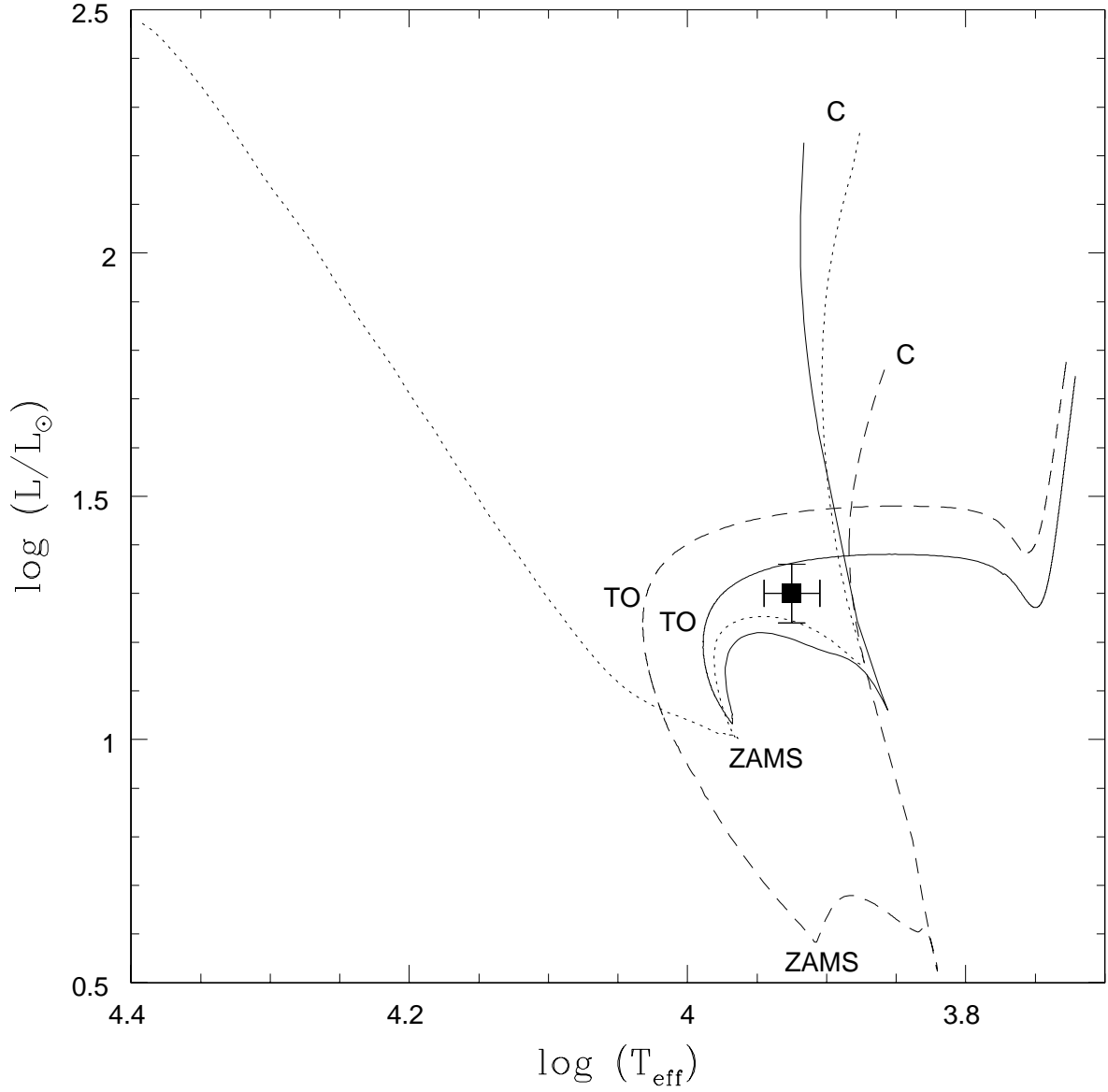}
\caption {Evolutionary tracks for case e (solid line), case f (dotted
line), and case k (dashed line), with the initial rotational velocity
divided by 5 in all cases. The different evolutionary points are
marked as in Figure \ref{caseKdiv}. The data point is BSS 19 in 47 Tuc
from \cite{SSL97}. \label{evolution}}
\end{figure}

The only blue straggler in a globular cluster to have its rotation
rate measured is BSS 19 in 47 Tucanae \citep{SSL97}. This star is
rotating at $v \sin i = 155 \pm 55$ km s$^{-1}$, which is consistent
with the rotation rates of the collision products presented in this
work.  The position of this star in the HR diagram is shown in Figure
\ref{evolution}. The rotation velocities of the collision products in
the vicinity of this star range from about 230 km s$^{-1}$ to 400 km
s$^{-1}$, which is consistent with the BSS 19 data point and a
reasonable inclination angle $i$ between $\sim20^\circ$ and
$\sim40^\circ$. Observations of more rotation rates of blue stragglers
in globular clusters are necessary to resolve some of the issues
involved in studying off-axis collision products.  However, we can say
that the conclusion of \cite{SSL97} that such a rapid rotation rate
necessarily indicates a binary merger formation for BSS 19, rather
than a collisional origin, is currently unwarranted. This conclusion
was based on the assumption of \cite{LL95} that collisional blue
stragglers would develop convective envelopes shortly after the
collision, and would then lose angular momentum through a magnetic
wind. As shown in Paper I and confirmed in this work, collision
products do not develop significant surface convection zones, and
therefore do not suffer from standard magnetic braking.

The other blue stragglers with measured rotation rates are found in
the old open cluster M67 \citep{PCL84}. These stars are rotating more
slowly than the blue straggler in 47 Tuc, and much more slowly than
the models presented here. A large fraction of these blue stragglers
are in binary systems. Could these two circumstances be related?  Blue
stragglers are thought to be formed through mass transfer in a binary
system as well as through direct stellar collisions.  Detailed
calculations will be necessary to determine if binary merger products
have different rotational properties than stellar collision products,
and if we can distinguish between the two blue straggler creation
mechanisms on the basis of their current rotation rates.

\acknowledgements This research was supported in part by NSF Grants
AST-9618116 and PHY-0070918, and NASA ATP Grant NAG5-8460 at MIT, and
by an award from Research Corporation and NSF Grant AST-0071165 at
Vassar.  This work was also supported by the National Computational
Science Alliance under grant AST980014N and utilized the NCSA SGI/Cray
Origin2000. F.A.R.\ was supported in part by an Alfred P.\ Sloan
Research Fellowship.  A.S.\ gratefully acknowledges support from the
Natural Sciences and Engineering Research Council of Canada.

\clearpage

\clearpage
\begin{figure}[ht!]
\caption {Snapshots of those SPH particles near the orbital plane
(i.e., within two smoothing lengths of $z=0$) for case e, a parabolic
collision between parent stars of masses $M_1=0.8M_\odot$ and
$M_2=0.6M_\odot$ at a pericenter separation $r_p=0.25(R_1+R_2)$.
Colors are used to indicate from which parent star the SPH particle
originated, as well as its initial entropy $A$: green and red points
are for particles originating in the $0.6M_\odot$ star, while yellow
and blue indicate particles from the $0.8M_\odot$; furthermore, green
and yellow correspond to particles with an entropy $A<10^{14}$ (cgs),
while red and blue correspond to particles with larger
entropies. (This figure can be downloaded from
http://www.astronomy.ohio-state.edu/~asills/papers/rotBS.html)
\label{color}
}
\end{figure}

\clearpage

\clearpage

\clearpage

\clearpage

\clearpage

\clearpage

\clearpage

\clearpage
\begin{deluxetable}{ccccccccccc}
\tabletypesize{\scriptsize}
\tablecaption{SUMMARY OF COLLISIONS \label{tbl-summary}}
\tablehead{
\colhead{Case} & \colhead{$M_1$}   & \colhead{$M_2$}   &
\colhead{$r_p$} & \colhead{$N$} & \colhead{$t_f$} & \colhead{$J$} &
\colhead{$T/|W|$} & \colhead{$M_r$}
& \colhead{$V_x$} & \colhead{$V_y$} \\
& \colhead{$[M_\odot]$}   & \colhead{$[M_\odot]$}   &
\colhead{$[R_1+R_2]$} &  & \colhead{[hours]} & \colhead{[g cm$^2$
s$^{-1}$]} &
& \colhead{$[M_\odot]$}
& \colhead{[km s$^{-1}$]} & \colhead{[km s$^{-1}$]} \\
(1)&(2)&(3)&(4)&(5)&(6)&(7)&(8)&(9)&(10)&(11)
}
\startdata
e & 0.8 & 0.6 & 0.25 & $1.05\times 10^5$ & 11.10 & $2.0\times 10^{51}$
& 0.101 & 1.359 & -1.8 & -2.7\\
f & 0.8 & 0.6 & 0.50 & $1.05\times 10^5$ & 24.64 & $2.8\times 10^{51}$
& 0.119 & 1.380 & -0.4 & -1.2\\
k & 0.6 & 0.6 & 0.25 & $9\times 10^4$ & 11.10 & $1.3\times 10^{51}$ &
0.085 & 1.158 & 0.0 & 0.0\\
j$^\prime$ & 0.6 & 0.6 & 0.01 & $1.6\times 10^3$ & 9.25 & $2.1\times
10^{50}$ & 0.005 & 1.147 & 0.0 & 0.0\\
\enddata
\end{deluxetable}

\begin{deluxetable}{ccccc}
\tablecaption{EVOLUTIONARY RESULTS \label{tbl-evolution}}
\tablehead{
\colhead{Case} &  \colhead{$V_{rot}$ (TO) } & \colhead{age at ZAMS} &
\colhead{age at TO} & \colhead {age at GB} \\
& (km s$^{-1}$) & (Gyr) & (Gyr) & (Gyr)
}
\startdata
e & 344 & 0.012 & 0.58 & 0.87 \\
f & 372 & 0.040 & 2.79 & -- \\
k & 353 & 0.077 & 4.19 & 4.38 \\
j$^\prime$  & 272 & 0.091 & 3.66 & 4.03 \\
\enddata
\end{deluxetable}

\end{document}